\documentclass{aastex}

\shorttitle{Globular Cluster NGC 7492}

\shortauthors{Lee et al.}

\begin{document}

\title{Mass Segregation and Tidal Tails of the Globular Cluster NGC 7492}

\author{Kang Hwan Lee}
\affil{Astrophysical Research Center for the Structure and
Evolution of the Cosmos, Sejong University, Seoul 143-747, Korea}
\affil{Centre for Astrophysics \& Planetary Science, School of
Physical Sciences, University of Kent, Canterbury, Kent CT2 7NR,
UK } \email{khlee@arcsec.sejong.ac.kr}

\author{Hyung Mok Lee}
\affil{Astronomy Program, SEES, Seoul National University, Seoul
151-742, Korea} \email{hmlee@astro.snu.ac.kr}

\author{Gregory G. Fahlman}
\affil{Herzberg Institute of Astrophysics, 5071 West Saanich Road,
Victoria, British Columbia V9E 2E7, Canada}
\email{Greg.Fahlman@nrc-cnrc.gc.ca}

\and

\author{Hwankyung Sung}
\affil{Department of Astronomy \& Space Science, Sejong
University, Seoul 143-747, Korea}
\email{sungh@arcsec.sejong.ac.kr}

\begin{abstract}

We present a wide field CCD photometric study of Galactic globular
cluster NGC 7492. The derived $VR$ color-magnitude diagram extends
down to about 3.5 mag below the cluster main sequence turn-off.
The field covers $42' \times 42'$ about 3 times larger than the
known tidal radius of this cluster. The sample of cluster member
candidates obtained by CMD-mask process has been used to construct
luminosity and mass functions and surface density map. NGC 7492
has a very flat mass function with very little variation in the
slope with the distance from the cluster center. However, there is
a clear evidence for the increase of the slope of the mass
function from inner to outer regions, indicating a mass
segregation of the cluster. The surface density map of NGC 7492
shows extensions toward the Galactic anticenter (northeast) and
northwest from the cluster center. A comparison of luminosity
function for stars in the tails with that for stars within the
tidal radius suggests that the extensions shown in the surface
density map could be a real feature. The overall shape of NGC 7492
is significantly flattened. If the flattened shape of the NGC 7492
is caused by its rotation, Galactic tidal field must have given
important influences, since the initial rotation would have been
almost completely removed by dynamical relaxation.

\end{abstract}


\section{INTRODUCTION}

Globular clusters have attracted considerable attentions for the
understanding of the process of Galaxy formation because they are
thought to be fossil relics from the early formation history of
the Galaxy. It is now well known that the globular clusters we see
today may be only the survivors of an initial population. From the
calculations of destruction rate, Lee \& Goodman (1995) and Gnedin
\& Ostriker (1997) predicted that possibly as many as half of the
present day Galactic globular clusters would be destroyed in the
next Hubble time. Two body relaxation makes the velocity
distribution of stars in a globular cluster toward a Maxwellian.
As some stars acquire enough energy to escape from the cluster,
the total mass slowly decreases (Spitzer \& Thuan 1972). The
evaporation is also accelerated by the energy equipartition
process which makes all stars have the same kinetic energy. It
causes the mass segregation through which high-mass stars sink
toward the central region of the cluster while low-mass stars have
higher velocities and tend to occupy the outer region of the
cluster (Spitzer 1987). Therefore mass segregation leads to the
preferential loss of low-mass stars. In addition, clusters
experience shocks when the external tidal field varies rapidly.
Passage through the Galactic disk and close to the Galactic bulge
also accelerate the destruction of clusters via gravitational
shocks.  Gnedin \& Ostriker (1997) found that tidal shocks
contribute at least as much as two body relaxation to the
destruction of the current globular clusters. The escaped stars
may remain in the vicinity of the cluster for several Galactic
orbits. As a result, the cluster is expected to have tidal tails.

Observational studies about tidal tails of globular clusters have
been carried out only recently because of difficulties in wide
field photometry. The signs of the existence of tidal tails around
globular clusters were found in many previous studies using radial
density profiles or two-dimensional density maps (Grillmair et al.
1995; Grillmair et al. 1996; Holland, Fahlman, \& Richer 1997;
Lehmann \& Scholz 1997; Leon, Bergond, \& Vallenari 1999; Leon,
Meylan, \& Combes 2000; Testa et al. 2000; Odenkirchen et al.
2001; Sohn et al. 2003; Lee et al. 2003; Odenkirchen et al. 2003).
The studies from radial density profiles suggest that many
clusters have weak halo or tails of unbound stars which might
result from tidal stripping. However, two-dimensional density maps
obtained in these studies did not clearly confirm this suggestion
because these are too complex and diffuse to be regarded as tidal
tails.

More convincing evidences for the tidal tails came out of deep and
wide CCD observations. Recently, using wide field photometric data
from the Sloan Digital Sky Survey (SDSS), Odenkirchen et al.
(2001, 2003) detected clear tidal tails around the Galactic
globular cluster Palomar 5, and showed that this cluster is being
tidally disrupted. The tails of Pal 5 extend over an arc of
$10^\circ$ on the sky, corresponding to a projected length of 4
kpc at the distance of the cluster. They also suggested that the
Pal 5 would be destroyed after the next disk crossing, which will
happen in about 100 Myr. From the wide field photometry using
CFH12K mosaic CCD, Lee et al. (2003) showed that the presence of
an extra-tidal profile extending out to at least $\sim 30'$ from
the center of Galactic globular cluster M92. Using CFH12K wide
field photometry, Sohn et al. (2003) found very weak tidal halos
around the remote young globular clusters Pal 3 and Pal 4.

In this paper we study the mass segregation effect and spatial
distribution of stars around the Galactic globular cluster NGC
7492 using the CHF12K $VR$ photometry data. NGC 7492 is one of the
most sparse globular clusters which is located far from the
Galactic center and plane ($R_{GC} = 24.9$ kpc, $Z = -23.1$ kpc)
and at 26.2 kpc from the sun. The fundamental parameters of NGC
7492 are listed in Table~1. The first CCD photometry for this
cluster was presented by Buonanno et al. (1987). They performed BV
CCD photometry for NGC 7492 down to $V \sim 23$ roughly 2 mag
fainter than the main-sequence turn-off. C\^{o}t\'{e}, Richer \&
Fahlman (1991, hearafter CRF91) carried out CCD photometry for
this cluster covering a field of $2.'2 \times 3.'5$ containing
cluster center. They presented a color-magnitude diagram (CMD) for
NGC 7492 that traces the main sequence to about one magnitude
fainter than that of Buonanno et al. (1987). They confirmed the
cluster metallicity to be [Fe/H] = $-1.51$, and derived a distance
modulus of $(m-M)_0 = 17.09 \pm 0.20$ mag by comparing the
fiducial isochrone of NGC 6752 given in VandenBerg, Bolte, \&
Stetson (1990). They also reported the discovery of 27 blue
straggler candidates which are found to be more centrally
concentrated than the cluster subgiants of similar brightness. If
the blue stragglers are more massive than the stars at the main
sequence turnoff, it would be the result of cluster mass
segregation. They also presented star counts for ten annuli
centered on the cluster core. However, they found no evidence for
mass segregation in either the surface density profiles or in the
luminosity functions at different radial positions. Possible
reasons of failure in finding mass segregation are that the
stellar mass range sampled by their data and/or the observed
region is too small to find the change of luminosity function (LF)
or mass function (MF). The small variation in the LF of NGC 7492
as a function of radius also can be expected from the study by
Pryor, Smith, \& McClure (1986). Recently Leon, Meylan, \& Combes
(2000) investigated the presence of tidal tails around 20 Galactic
globular clusters including NGC 7492. Except for a tiny extension
pointing toward the Galactic center, they did not find any other
signs of tidal tail in NGC 7492. In this study, we present the
investigation of mass segregation and tidal tails of NGC 7492
using deeper and wider field CCD photometry than that of CRF91.

\placetable{tab1}

In the following section, we present the observation and data
reduction process. In section 3, we show CMDs of NGC 7492 and
describe CMD-mask algorithm which select the sample of cluster
member candidates. In section 4, we examine the mass segregation
effect of the cluster using luminosity and mass functions of the
cluster. In section 5, we examine the spatial distribution of
stars around the cluster using surface density map and luminosity
functions. Final results are summarized in section 6.

\section{OBSERVATION AND DATA REDUCTION}

The observations were made with the 3.6 m Canada-France-Hawaii
Telescope (CFHT) during 1999 Oct. 17-18 using CFH12K, a $6 \times
2$ mosaic of $2048\times 4096$ CCDs. The CFH12K has an angular
scale of $0.''206$/pixel at the $f/4$ prime focus of the CFHT,
giving a field of view of $42' \times 28'$. Each chip covers an
area of $7' \times 14'$. Two adjacent fields were observed using
$V$ and $R$ filters, covering an area of $42' \times 42'$, which
extends beyond 3 times larger than the known tidal radius of this
cluster, $8.'35$ (Harris 1996). The positions of the observed
fields relative to the center of NGC 7492 are shown in Figure~1,
and observational information is given in Table~2. The points in
Figure~1 represent the stars brighter than $V$ = 24 and the point
size is proportional to the brightness of stars. Each field was
properly dithered to fill in the gaps between chips. We also
obtained frames of fields containing Landolt (1992) standard stars
for photometric calibration. All of the science images were
obtained under good seeing conditions (FWHM of $\sim 0.''8$).
Several twilight flat-field, bias and dark frames were also taken.

\placefigure{fig1}

\placetable{tab2}

Pre-processing of the raw data, bias and dark subtraction and flat
fielding, was done using the FITS Large Images Processing Software
(FLIPS), a very efficient package  for the reduction of CCD mosaic
images developed by Jean-Charles Cuillandre at CFHT (Kalirai et
al. 2001). FLIPS is designed to operate on individual chips within
the CFH12K mosaic. First, the good exposures for each of the bias
and dark exposures were median combined. For the flat-field images
of each filter we averaged and sigma clipped the flats taken from
observing run. Median combined bias and dark frames were then
subtracted from the flat-field and object images. After the object
images were debiased, the images were flat-fielded to account for
pixel-to-pixel variations. FLIPS normalizes the background sky
values to the chip with the highest sky level (lowest gain) and
provides for a scaled data set with a smooth background on all
chips. This makes the instrumental zero points for each chip
almost equal. The statistics at various positions in the mosaic
show that the flat-fielding to be typically better than 1\% in
both $V$ and $R$ filters.

Instrumental magnitudes of the point sources in the images were
derived using the programs DAOPHOT II/ALLSTAR (Stetson 1994). For
long exposure data (600s) we averaged the images of each filter
using the ALIGN and IMCOMBRED commands within FLIPS. Short
exposure images were used to get the instrumental magnitudes for
stars saturated in long exposure images. The analysis is done
separately for each chip because the point-spread function (PSF)
is different for each of the 12 CCDs on the mosaic. We used a
variable PSF to account for small changes in the profiles of stars
over the large range of each CCD. We used the stellarity index of
SExtractor (Bertin \& Arnouts 1996) for separating stars from
background galaxies.
As an example, in Figure~2 we show this parameter for sources
detected in F1. Those objects with a stellarity index greater than
0.8 were considered to be stars. We applied a little more strict
cut to avoid overestimation of faint stars which could exaggerate
the mass segregation effect and tidal extensions of the cluster.
This process also eliminates bad pixels caused by cosmic-ray hits.
After further cuts, based on the two DAOPHOT parameters, the
magnitude error ($< 0.2$) and the PSF fitting parameter, $\chi^2$
($< 2$), (we didn't consider sharpness parameter because we
already used stellarity index) were made to eliminate spurious
detections, the data sets in each filter were matched and the
positions, instrumental magnitudes, and colors were obtained for
all the stars.

\placefigure{fig2}

A number of Landolt (1992) standard stars in SA-92 were obtained
during the observing run in order to convert the instrumental
magnitudes to standard photometric magnitudes. The transformation
equations to derive standard magnitudes are as follows;
\[
V = v + \alpha X + \beta (V - R) + Z_v,
\]
\[
R = r + \alpha' X + \beta' (V - R) + Z_r.
\]
In these equations $v$ and $r$ are instrumental magnitudes and
$\alpha$, $\alpha'$ are the extinction coefficients of the airmass
term $X$, $\beta$, $\beta'$ are the transformation coefficient of
the color term $(V-R)$, and $Z_v$, $Z_r$ are the zero-point shift
for the $V$ and $R$ bands. Since we could get only $1 \sim 3$
standard stars on each chip, we applied the standard values for
the color term and the atmospheric extinction coefficient given on
the CFHT home page. Separate calibrations are actually required
for each chip, because there are systematic differences between
chips. However, the color term does not significantly change with
different exposures or nights of observations and the extinction
coefficient is very stable at CFHT site (Kalirai et al. 2001). We
used standard stars to get the zero points of each chip which are
the most critical part of the transformation equations. The
instrumental zero points for the chips were almost identical
because FLIPS normalizes the background sky value. The mean value
of the differences between our photometry of the Landolt (1992)
standard stars and the standard values is 0.01. To check the
accuracy of photometry and calibration we used stars in the
dithered frames in each field and overlapped region between the
two fields. The same stars in the different chips and at different
locations in the field of view of the instrument showed
differences less than limiting magnitude error (0.2) in all chips.
Figure~3 shows the magnitude differences between the same stars in
the different chips.

\placefigure{fig3}

\section{COLOR-MAGNITUDE DIAGRAM}

The derived CMD based on the long and short exposure data for the
entire observed region of the cluster is shown in Figure~4.
Well-defined cluster sequence as well as field stars scattered
over the entire CMD can be easily seen. In Figure~4, we can see a
well-defined blue horizontal branch (BHB), and the absence of a
red horizontal branch (RHB). Also clearly visible are the red
giant branch (RGB), subgiant branch (SGB), the main sequence (MS)
turnoff at $V \sim 21$ mag, and a MS that is extended down to
about $V \sim 24.5$ mag which is roughly 3.5 mag fainter than
turnoff. The characteristics of CMD morphology are described by
CRF91 in detail. Since our main concern is the distribution of
cluster stars, the contamination from field stars needs to be
removed.

\placefigure{fig4}

To select cluster stars, we used the technique of the CMD-mask
algorithm, which was introduced by Grillmair et al. (1995). In
this way we can differentiate cluster members from the background
and foreground field stars by means of comparing stars inside the
tidal radius of the cluster with the field stars in a distant part
of the field. The filtering mask is empirically chosen so as to
optimize the ratio of cluster stars to field stars in the
relatively sparsely populated outer region of the cluster.

We regarded the regions beyond 2.5 times of tidal radius ($r
> 20.'9$) as background regions which is shown in Figure~1. In
Figure~5, we show CMDs of the cluster region inside the tidal
radius($r < r_t = 8.'35$), and outer region ($r
> 2.5 \times r_t = 20.'9$). The procedure was performed in the
region $-0.6 < V-R < 0.9$ and $15 < V < 25.5$ which is indicated
by rectangular boxes in Figure~5. The region was subdivided into a
$30 \times 55$ array in which the individual subgrid elements were
0.05 mag wide in $V-R$ and 0.2 mag high in $V$.

\placefigure{fig5}

Assuming that the color-magnitude distribution of the field stars
is constant across the observed field, a color-magnitude sequence
for the cluster can be estimated from (Grillmair et al. 1995):
\[
f_{cl}(i,j) = n_{cl}(i,j) - gn_f (i,j),
\]
where $n_{cl}(i,j)$ and $n_f (i,j)$ refer to the number of stars
with color index $i$ and magnitude index $j$, counted in an area
of circle with radius $r< r_t$ around the cluster center, and
around the background field region, respectively. The factor $g$
is the ratio of the area of the cluster region to that of the
background field region. The signal-to-noise ratio of the expected
true number of cluster stars for each subarea was computed from:
\[
s(i,j) = {f_{cl}(i,j) \over {\sqrt{n_{cl}(i,j) + g^2 n_f (i,j)}}}.
\]

From the signal-to-noise $s$ we obtain a filtering mask by
isolating the region in the color-magnitude subgrid with $s >
s_{lim}$. To select the optimal range of color and magnitude, the
elements of $s(i,j)$ were sorted into a gradually descending order
over the one-dimensional index $l$ first. From the subgrid element
with the highest $s$ value, cumulative star counts were carried
out in the cluster region using progressively larger areas of the
CMD, $a_k = ka_l $, where $a_l = 0.016$mag$^2$ is the area of a
single element in the color-magnitude array. Then the cumulative
signal-to-noise ratio, $S(a_k )$, can be computed from:
\[
S(a_k ) = {N_{cl}(a_k ) - gN_f (a_k ) \over {\sqrt{N_{cl}(a_k ) +
g^2 N_f (a_k )}}},
\]
where $N_{cl}(a_k )$ and $N_f (a_k )$ are the cumulative number of
stars in the corresponding subarea in the cluster region and in
the outer background field region, respectively. Now, $n_{cl}(l)$
and $n_f (l)$ refer to the number of stars within the cluster
region and background field region, having ordered color-magnitude
index $l$. This cumulative function reaches a maximum for a
particular subarea of the CMD plane. Based on the peak value of
$S$, a threshold value of $s_{lim}$ is determined. The heavy lines
in Figure~5 indicate the filtering mask differentiating cluster
members from the field stars. By extracting stars outside the
filtering mask in the CMD, the field contamination was reduced by
a factor of $\sim 4$.

\section{MASS SEGREGATION}

It is well known that the radial variation of luminosity functions
(LFs) and mass functions (MFs) give a crucial hint of the mass
segregation effect in a globular cluster. LFs were constructed
using all CMD-selected stars within the cluster's tidal radius. To
build a reliable LF especially in the crowded field,
incompleteness corrections must be applied correctly. To correct
for the incompleteness of our photometry, we ran artificial star
tests on the $V$ frames, using ADDSTAR routine in IRAF/DAOPHOT.
First, we added artificial stars in each 0.5 mag bin randomly on
all original images. The number of added stars was designed not to
exceed 10\% of the total number of stars that is actually present
in that bin so as not to enhance original crowding. The new frames
obtained in this way were then reduced in an identical manner used
on the original frame. Figure~6 shows a sample of the plots of the
differences between input and output magnitudes for recovered
artificial stars selected by the same criteria applied to original
image reduction. Since most of values do not exceed 0.3 mag, the
stars which have magnitude differences smaller than 0.3 mag are
considered as recovered stars. We repeated these procedures 10
times on each chip to obtain meaningful statistics in each
magnitude bin. Finally, the incompleteness correction factor $f$
is obtained by $f = n_{rec}/n_{add}$, where $n_{add}$ is the
number of added stars and $n_{rec}$ is the number of recovered
stars. To examine the accuracy of the incompleteness correction,
we ran artificial star tests in another way. We selected two chips
in each field of $V$ and $R$ frames. Then the artificial stars
which have similar color band of the cluster MS and RGB stars have
been added on the $V$ and $R$ images with the same method
described above. The new frames were then reduced in an identical
manner used on the original frame. The stars, which have magnitude
differences smaller than 0.3 mag and color differences smaller
than 0.2 mag are considered as recovered stars. The incompleteness
correction factors obtained from this way were in accord with the
previous results within 5\% on the same chips. The incompleteness
correction factors depended on the distance from the cluster
center and did not show any differences between chips. The
resulting incompleteness correction factors were then applied to
the raw LFs. To investigate any spatial variation of LF, the LFs
for the inner region ($0' < r <1.'3$) and the outer region ($1.'3
< r < 8.'35$) are constructed separately. The regions are divided
to contain similar number of stars in each region. The corrected
number of stars in each magnitude bin and the incompleteness
factors are given in Table~3, and the corrected LFs are plotted in
Figure~7. The histogram in Figure~7 is the LF for all selected
stars within tidal radius ($r < 8.'35$). The solid line overlaid
on the inner LF is that of NGC 7492 for $0' < r < 2'$ derived by
CRF91, which agrees very well with ours. This agreement implies
that the incompleteness corrections are properly applied.

\placefigure{fig6}

\placetable{tab3}

The presence of mass segregation be seen from the radial variation
of LFs. The slope of LFs of globular clusters are known to
increase toward the outer parts of the cluster where low mass
stars preferentially occupy as a result of mass segregation.
However, we could not find any definite evidence from Figure~7
alone. As discussed below, the small variation in the LF of NGC
7492 as a function of radius is not an unexpected result. The LFs
were converted into mass functions (MFs) using the mass-luminosity
relation of Baraffe et al. (1997) and a distance modulus of
$(m-M)_0 = 17.09$ (CRF91). The calculated MFs are shown in
Figure~8. Due to the small number of stars brighter than the MS
turn-off, only stars with $V > 20$ mag have been used. Filled
circles represent the MF for the inner region ($0' < r < 1.'3$)
while open circles for outer region ($1.'3 < r < 8.'35$). Open
triangles represent the MF derived by CRF91. Because of our LF
extends only to $M_V \sim 7.5$ the MF is defined only in the range
$0.6 \sim 0.8 M_\odot$. For this reason, it is difficult to say
anything definitely about the slope of the main sequence MF based
solely on Figure~8. CRF91 derived a slope of the MF $x = -1.1\pm
0.5$ using the theoretical LF from Drukier et al. (1988). Here $x$
represents the slope of a MF of the form
\[
\Phi (M) dM \propto M^{-(1+x)} dM.
\]
We insert lines in Figure~8 which show several power law indices
covering the slope from CRF91. Although it is difficult to fit a
single power law to MFs of this cluster, we can conclude that the
$x$ value lies within the range from previous study. It is
noteworthy that the cluster has a very flat MF and the variation
of the MFs from inner to outer regions is very small. Although the
slope changes by very small amount, we can see a clear tendency of
increase in the slope of MF from inner to outer regions in
Figure~8. It could be a sign of mass segregation in this cluster.

\placefigure{fig7}

\placefigure{fig8}

Small change in the MF slope does not mean that this cluster is
not fully relaxed. Pryor, Smith \& McClure (1986) have
investigated the effects of mass segregation on main sequence MFs
for globular clusters with various central concentrations. They
found that mass segregation depends on core concentration
parameter ($c$) and the slope of mass functions ($x$). They showed
that the mass segregation effect appears more clearly in models
with large concentrations and steep (larger $x$) mass functions.
For a fully relaxed cluster having $c = 1.0$ and $x \sim -1$,
there will be very little change in the MFs due to mass
segregation. Using the HST photometry, Grillmair \& Smith (2001)
investigated the main-sequence LF of Pal 5. According to their
results, Pal 5 has a very flat MF ($x = -0.5$) and shows very
little evidence for mass differentiation between the core of the
cluster and the half-mass radius. They suggested that Pal 5 has
lost a large fraction of its original stellar content as a result
of tidal shocking. As mentioned in section 1, Pal 5 appears to be
in the final phase of tidal disruption (Odenkirchen et al. 2003).
In any globular cluster the form of the present day mass function
depends on the initial mass function and the process of dynamical
evolution of the cluster. McClure et al. (1986) suggested that the
slope of MF is related with metallicity of cluster. They proposed
that the observed strong dependence of MF slopes on metallicity
reflects, at least in part, properties of the initial mass
functions with which the clusters formed. Later, Djorgovski,
Piotto, \& Capaccioli (1993) demonstrated that the MF slopes are
determined not only by the metallicity but also by the location in
the Galaxy. They showed that the clusters closer to the Galactic
center have flatter MFs. At a given Galactocentric distance
($R_{GC}$), clusters with a smaller distance from the Galactic
plane ($Z$) have flatter MFs, and at a given position, clusters
with lower metallicity have steeper MFs. From N-body simulations,
Capaccioli, Piotto, \& Stiavelli (1993) interpreted the dependence
on position as the effect of tidal shocks. Disk and bulge shocking
acts preferentially on stars located in the outer region of the
cluster and leads to a loss of low mass stars. As a result the
clusters near the Galactic center and plane, tend to have flatter
MFs. Since NGC 7492 has intermediate metallicity ([Fe/H] =
$-1.51$) among the Galactic globular clusters and large distant
from Galactic center and plane ($R_{GC} = 24.9$ kpc; $Z = -23.1$
kpc), flat MF of NGC 7492 can not be interpreted neither by its
metallicity nor by Galactic position. However, dynamical evolution
of a globular cluster can not be determined only by present
position but also by cluster's orbit. Piotto, Cool, \& King (1997)
presented a comparison of deep HST LFs of four Galactic globular
clusters. They showed that three of four clusters (M15, M30, and
M92) have nearly identical LFs, whereas NGC 6397 has a distinctly
different LF, especially in the fainter part. They suggested that
the three globular clusters which have similar LFs were formed
with similar MFs and suffered very little changes (or experienced
similar changes regardless of the location), and that the
deficiency of low-mass stars in NGC 6397 was due to tidal shocks,
stellar evaporation through internal relaxation, or a combination
of the two. Relatively flat MF of NGC 6397 could be explained by
its orbit given by Dauphle et al. (1996) which is very vulnerable
to tidal shocks. Later, comparing MF slopes for 7 Galactic
globular clusters Piotto \& Zoccali (1999) suggested that the
flattening of MF slopes might be related to the cluster's
dynamical evolution. From the N-body simulations, Baumgardt \&
Makino (2003) predicted that the slope of MFs of the clusters
should decrease constantly as the clusters evolve, and there is a
good correlation between the slopes and the cluster lifetimes. If
the flat MF of NGC 7492 is not primordial it might be a result of
dynamical evolution of this cluster. When the Galactic orbit of
NGC 7492 is determined the influence of Galactic tidal shocks on
this cluster could be known.

\section{SPATIAL DISTRIBUTION OF STARS}

\subsection{Surface Density Map}

In order to characterize the distribution of stars around the
cluster, we constructed surface number density map. The sample of
cluster member candidates obtained by CMD-mask process in the
previous section was used to construct surface density map.
Initially, the candidate cluster stars in the surveyed region were
binned on a grid of $0.'2 \times 0.'2$. Unfortunately, our
surveyed region is not wide enough to get proper background maps.
However, we used field stars having color greater than 0.9 to
check the gradient of distribution of field stars and possible
photometric biases in the entire surveyed region. After
constructing the map of field stars by binning $1' \times 1'$, we
smoothed it to get a surface map dominated only by the smooth
gradient of field stars. We subtracted this field star map from
the CMD-selected one. We then convolved the map with a Gaussian
kernel of width $1.'0$. The resulting smoothed surface density map
is shown in Figure~9.

\placefigure{fig9}

We overlaid contour levels and marked the tidal radius with a
thick circle. The arrow indicates the direction to the galactic
center. We can see a marginal extension toward the Galactic
anticenter (northeast) in Figure~9. Also there seems to be a small
extension to the northwest from the cluster center. Many N-body
simulations of globular cluster tidal tails show that the stars
evaporated from the cluster form a twisted, two-sided lobe
distribution of cluster member stars (Combes, Leon, \& Meylan
1999; Yim \& Lee 2002; Lee et al. 2004). However, the shape of
such lobes depends on the orientation of the cluster's orbit to
our line of sight. Combes, Leon, \& Meylan (1999) showed that the
lobes can be asymmetric and even seen one-sided according to the
cluster's orbits and projection effects in their Figure~13 and
Figure~15. The extensions shown in Figure~9 can not be regarded as
a real feature solely based on the surface density map. We will
check the significance of the extensions using luminosity
functions in the following subsection.

From Figure~9, we can see that the cluster NGC 7492 has very
flattend shape. White \& Shawl (1987) have measured the projected
axial ratio $(b/a)$ of 100 Galactic globular clusters, where $a$
and $b$ denote semi-major and semi-minor radii, respectively. They
obtained the value of $<b/a> = 0.76$ from NGC 7492 while mean
axial ratio was $<b/a> = 0.93 \pm 0.01$. We used ELLIPSE task in
IRAF/STSDAS to fit the elliptical shape of the cluster. Figure~10
shows the axial ratio of the cluster as a function of the
semi-major axes of the ellipses. The axial ratio decreased and
reached to a value of 0.77.  White \& Shawl (1987) argued that the
flattened shape of the globular clusters can be caused by either
anisotropy in velocity dispersion or rotation. The tidally
truncated clusters tends to become quickly isotropic (Takahashi \&
Lee 2000), the flattening is likely due to rotation (Combes, Leon
\& Meylan 1999). Kinematical data also showed that the flattening
could be explained by rotation, and that the minor axes are nearly
coincident with the rotation axes (Meylan \& Mayor 1986). The
initial angular momentum of a cluster is expected to disappear
with time by the escape of stars carrying some angular momentum.
Kim et al. (2002) showed using the Fokker-Plank method that the
initial rotation decreases with time after core-collapse and
finally disappear. As a result, the flattening of a globular
cluster could be related with its dynamical age. From a
compilation of data on globular cluster flattenings in our Galaxy
and M31, Davoust \& Prugniel (1990) showed that globular clusters
with shorter relaxation times tend to be rounder. They suggested a
scenario which globular clusters are initially flattened and
become rounder as they lose stars and angular momentum. According
to this scenario, NGC 7492 would be the one which has the youngest
dynamical age among the Galactic globular clusters. However
considering the flat MF of NGC 7492, it is unlikely that this
cluster maintain initial dynamical status. If the flattening of
NGC 7492 cannot be explained only by initial rotation of this
cluster, we should consider some alternative sources of globular
cluster rotation. One possible alternative is tidal interaction
with the Galaxy. From N-body simulations Lee et al. (2004) showed
that the globular clusters gain angular momentum from tidal
interactions with host galaxy. If a cluster rotates by the tidal
interaction with the galaxy, the orientation of the rotation axis
would be the same as its revolution axis. When we get the orbit of
NGC 7492, the influence of tidal interaction on its rotation could
be examined.

\placefigure{fig10}

\subsection{Luminosity Functions}

Figure~9 shows marginal extensions of stellar distributions around
NGC 7492 toward northeast and northwest from the cluster center.
However we can not regard the extensions shown in Figure~9 as a
real feature using only the surface density map. We therefore
analyzed the LFs of the stars around the clusters to check the
significance of the extensions.

In figure~11, we show the cumulative LF for stars (solid line)
located in outer northern part ($r > r_t$) where extensions
appear. The dashed line indicates the LF for stars within the
tidal radius ($r_t$) of NGC 7492. For comparison, we also show the
LF for stars in southern part of the cluster as a dotted line.
These LFs are constructed using CMD-selected stars. Incompleteness
corrections are properly applied to each LF as described in the
previous section. The LFs are normalized by the total number of
stars for each selected group. A Kolmogorov-Smirnov (K-S) test of
the LFs for stars in the outer northern part (solid line) compared
to that for stars within the tidal radius (dashed line) gives a
significance level of 75\% probability that both cumulative LFs
follow the same distribution. On the other hand, the K-S
significance level of cumulative LF for stars in the southern part
(dotted line) is less than 1\%. Since mass segregation effect of
this cluster is very small, we can expect that the LF for stars in
the tidal tails of the cluster would be almost the same with that
for stars within the tidal radius in the limit of our analysis
(Grillmair \& Smith 2001). Odenkirchen (2003) also showed that the
LF for the stars in the tidal tails of Pal 5 is in very good
agreement with the LF in the cluster. Theses indicate that the
extensions shown in the surface density map (Figure~9) could be a
real feature.

\placefigure{fig11}

\section{SUMMARY}

We have carried out a wide field CCD photometry of NGC 7492 and
investigated the dynamical structure of the cluster. The
observations with the CFH12K mosaic CCD cover the area of $42'
\times 42'$, about 3 times larger than the known tidal radius of
this cluster. We used the technique of the CMD-mask algorithm to
select cluster member star candidates, and we have used these
stars to examine the characteristics of the spatial distribution
of stars around the clusters. The mass segregation effect and
stellar distribution around the cluster have been investigated by
means of a completeness-corrected LFs and MFs and the surface
density map.

The slope of cluster main sequence MF slightly increases toward
the outer part of the cluster, as expected from mass segregation.
It could be a verification of mass segregation effect of this
cluster which was suggested doubtfully only from the distribution
of blue stragglers by CRF91. Although the change in the MF slope
is very small it does not mean that this cluster is not fully
relaxed. The relatively flat MF slope of NGC 7492 and its small
mass might be a result of dynamical evolution of this cluster.

Surface density map shows possible extensions of spatial stellar
distributions beyond the tidal radius of NGC 7492. The extensions
of the tails of NGC 7492 are oriented toward the Galactic
anticenter (northeast) and northwest from the cluster center. A
comparison of LF for stars in the tails with that for stars within
the tidal radius suggests that the extensions shown in the surface
density map could be a real feature. Surface density map also
shows that the NGC 7492 has very flat shape. The axial ratio
($b/a$) of this cluster decrease with semi-major axis and reached
to a value of 0.77 which is one of the smallest values the
Galactic globular clusters have. If the flattened shape of this
cluster is caused by its rotation, it might have been much
affected by Galactic tidal field.

Pryor et al. (1991) argued that low concentration (low-$c$)
clusters lost much of their original stellar mass over extended
periods of time through the evaporation and stripping of stars.
Recently, Odenkirchen et al. (2003) showed clear evidences that
the low-mass, low-$c$ Galactic halo cluster Pal 5 is being tidally
disrupted. In addition to previous studies, flat mass function,
tidal extensions, and flat shape of the cluster which are
investigated in this study suggest that NGC 7492 might have
experienced a lot of dynamical processes through tidal interaction
with the Galaxy. However, kinematic data of NGC 7492 are not
obtained yet. The information on the orbits of individual halo
clusters can produce powerful constraints on the Galactic
potential. We need spectroscopic observations and proper motion
studies about NGC 7492 to get the information about the rotation
and the Galactic orbit of the cluster. We also encourage deeper
and wider CCD photometry of this cluster to find possible tidal
tails extending over the coverage of this study.

HML was supported by the KOSEF grant No. R01-1998-00023. H.S.
acknowledges the support of the Korea Science and Engineering
Foundation (KOSEF) to the Astrophysical Research Center for the
Structure and Evolution of the Cosmos (ARCSEC$''$) at Sejong
University.

\clearpage

\figcaption[Lee.fig01.ps]{Location of 2 observed fields, with the
origin set on the cluster center. The circles indicate known tidal
radius $r_t = 8.'35$ (inner) and 2.5 times of tidal radius ($r =
20.'9$). Each frame has a field of view of $42' \times 28'$. The
points represent the stars brighter than $V = 24$. The point size
is proportional to the brightness of stars. \label{fig1}}

\figcaption[Lee.fig02.eps]{The stellarity index of objects in F1
region. Sources with a stellarity index greater than 0.8 are
considered to be stars. \label{fig2}}

\figcaption[Lee.fig03.eps]{The differences between the same stars
in the different chips. \label{fig3}}

\figcaption[Lee.fig04.eps]{CMD for all calibrated stars. This is
the result of merging the long and short exposure data.
\label{fig4}}

\figcaption[Lee.fig05.eps]{CMDs for the stars within the tidal
radius ($r < r_t = 8.'35$), and outer region ($r > 20.'9$). The
heavy lines indicate criteria for selecting cluster member
candidates to reduce field star contribution. \label{fig5}}

\figcaption[Lee.fig06.eps]{The result of artificial star test for
one direction from cluster center. The difference $\Delta$ V is
$V_{in} - V_{out}$. \label{fig6}}

\figcaption[Lee.fig07.eps]{Luminosity functions of inner ($0' < r
<1.'3$) and outer ($1.'3 < r < 8.'35$) region for NGC 7492. The
histogram is the LF for all stars within $r < 8.'35$. A solid line
overlayed with inner LF is the result from C\^{o}t\'{e}, Richer \&
Fahlman (1991). For convenient comparison, the LFs are arbitrarily
shifted. \label{fig7}}

\figcaption[Lee.fig8.eps]{Mass functions of inner ($0' < r <1.'3$)
and outer ($1.'3 < r < 8.'35$) region for NGC 7492. Filled circles
represent the MF for the inner region while open circles for outer
region. Open triangles in the upper panel are the result from
CRF91. The solid lines represent several power law indices.
\label{fig8}}

\figcaption[Lee.fig9.eps]{Surface density map and contours of
levels of and all selected stars. The tidal radius is marked as a
thick circle. The arrow indicates the direction of the Galactic
center. Contours are drawn at 1, 2, 3 and 5 $\sigma$ of the
background. \label{fig9}}

\figcaption[Lee.fig10.eps]{The axial ratio ($b/a$) of the cluster
NGC 7492 as a function of the length of semi-major radius.
\label{fig10}}

\figcaption[Lee.fig11.eps]{The cumulative LFs for stars located in
the northern part where tidal extensions are shown (solid line)
and within the tidal radius (dashed line) of the cluster NGC 7492.
Dotted line indicates the LF for southern part of outer region of
the cluster. \label{fig11}}

\clearpage

\setcounter{table}{0}
\begin{deluxetable}{lrrrrrrrr}
\setlength{\tabcolsep}{3.5mm} \tablecaption{Coordinates and
Physical Parameters for NGC 7492\label{tab1}} \tablewidth{0pt}
\tablehead{\colhead{Parameter} & \colhead{Value} &
\colhead{Reference}}

\startdata
$\alpha (2000)$ & $23^h 08^m 26.7^s$  & Harris (1996) \\
$\delta (2000)$ & $-15^o 36' 41''$ & Harris (1996) \\
$l(2000)$(deg) & 53.39 & Harris (1996) \\
$b(2000)$(deg) & -63.48 & Harris (1996) \\
log $\rho_0$ ($M_\odot$ /pc$^3$) & 0.97 &  Harris (1996) \\
$r_t$(arcmin) & 8.35 &  Harris (1996) \\
$c$ = log $(r_t /r_c )$ & 1.00 & Harris (1996) \\
\ [Fe/H] & -1.51 & CRF91 \\
$(m - M)_0$ & 17.09 & CRF91 \\
$E(B - V)$ & 0.00 & CRF91 \\
\enddata
\end{deluxetable}

\clearpage

\setcounter{table}{1}
\begin{deluxetable}{lrrrrrrrr}
\setlength{\tabcolsep}{3.5mm} \tablecaption{Obsevational
Information of NGC 7492 \label{tab2}} \tablewidth{0pt} \tablehead{
\colhead{Field} & \colhead{$V$-Filter} & \colhead{date} &
\colhead{$R$-Filter} & \colhead{date} }

\startdata

F1 & $2 \times 20$s & Oct. 17 1999 & $1 \times 20$s & Oct. 17 1999 \\
F1 & $1 \times 100$s & Oct. 17 1999 & $1 \times 100$s & Oct. 17 1999 \\
F1 & $6 \times 600$s & Oct. 17 1999 & $6 \times 600$s & Oct. 17 1999 \\
F2 & $1 \times 20$s & Oct. 18 1999 & $1 \times 20$s & Oct. 18 1999 \\
F2 & $1 \times 100$s & Oct. 18 1999 & $1 \times 100$s & Oct. 18 1999 \\
F2 & $4 \times 600$s & Oct. 18 1999 & $4 \times 600$s & Oct. 18 1999 \\

\enddata
\end{deluxetable}

\clearpage

\setcounter{table}{2}
\begin{deluxetable}{lrrrrrrrr}
\setlength{\tabcolsep}{3.5mm} \tablecaption{Corrected Luminosity
Functions and Incompleteness Correction Factors \label{tab3}}
\tablewidth{0pt} \tablehead{ \colhead{V} &
\multicolumn{2}{c}{Inner ($0'< r < 1.'3$)} &
\multicolumn{2}{c}{Outer ($1.'3 < r < 8'.35$)} &
\multicolumn{2}{c}{Global} & \\
\colhead{} & \colhead{N} & \colhead{$f$} & \colhead{N} &
\colhead{$f$} & \colhead{N} & \colhead{$f$} }

\startdata

15.5-16.5 & 4.0 & 1.0 & 3.0 & 1.0 & 7.0 & 1.0 \\
16.5-17.5 & 10.0 & 1.0 & 12.0 & 1.0 & 22.0 & 1.0 \\
17.5-18.5 & 32.0 & 1.0 & 26.0 & 1.0 & 58.0 & 1.0 \\
18.5-19.5 & 28.0 & 1.0 & 22.0 & 1.0 & 50.0 & 1.0 \\
19.5-20.5 & 73.2 & 0.99 & 49.5 & 0.99 & 122.7 & 0.98 \\
20.5-21.5 & 490.0 & 0.90 & 310.4 & 0.96 & 800.4 & 0.94 \\
21.5-22.5 & 963.3 & 0.79 & 669.2 & 0.91 & 1632.5 & 0.87 \\
22.5-23.5 & 1263.8 & 0.69 & 943.3 & 0.90 & 2207.1 & 0.80 \\
23.5-24.5 & 1274.4 & 0.43 & 1068.6 & 0.70 & 2343.0 & 0.58 \\

\enddata
\end{deluxetable}

\clearpage

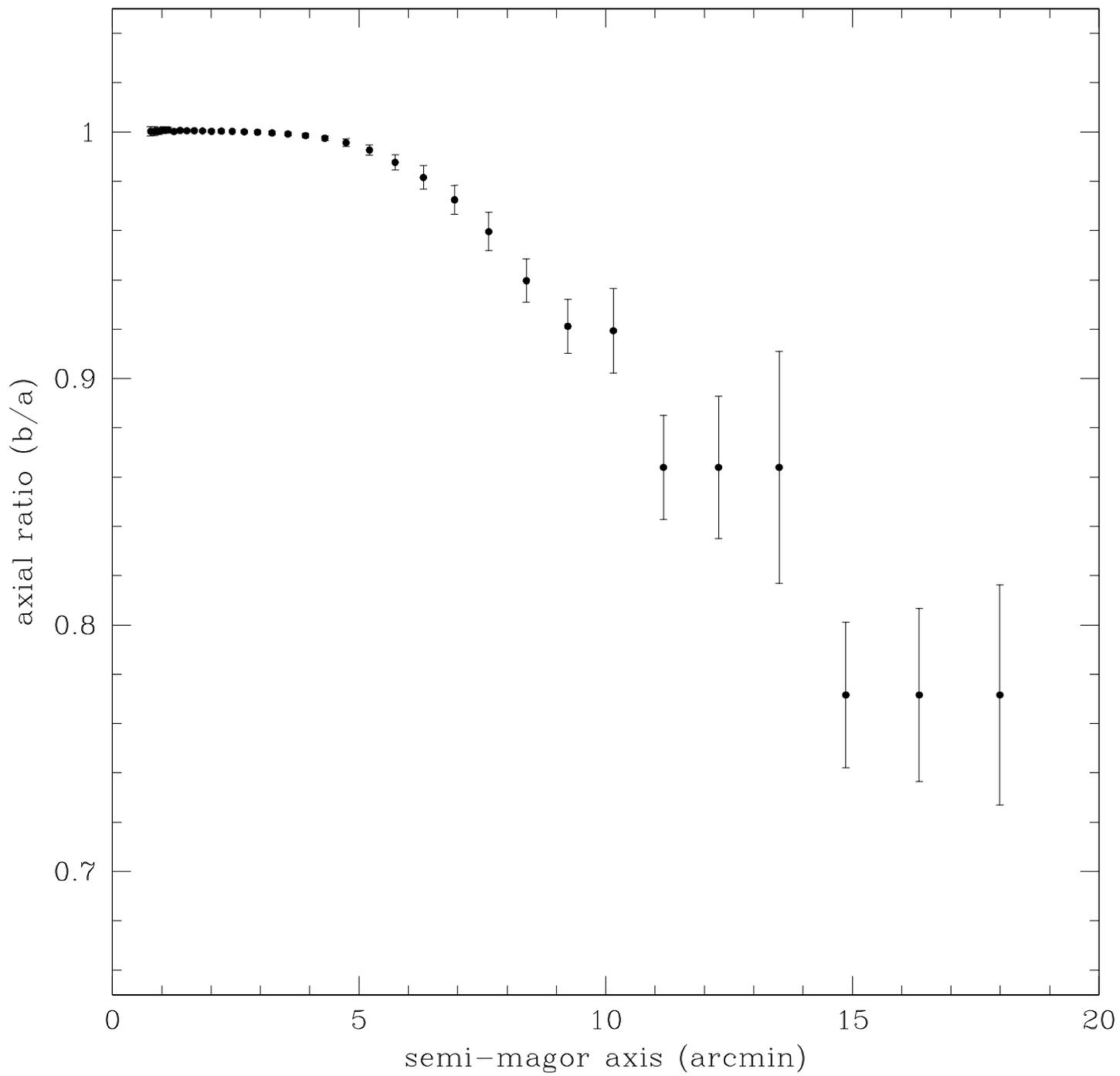
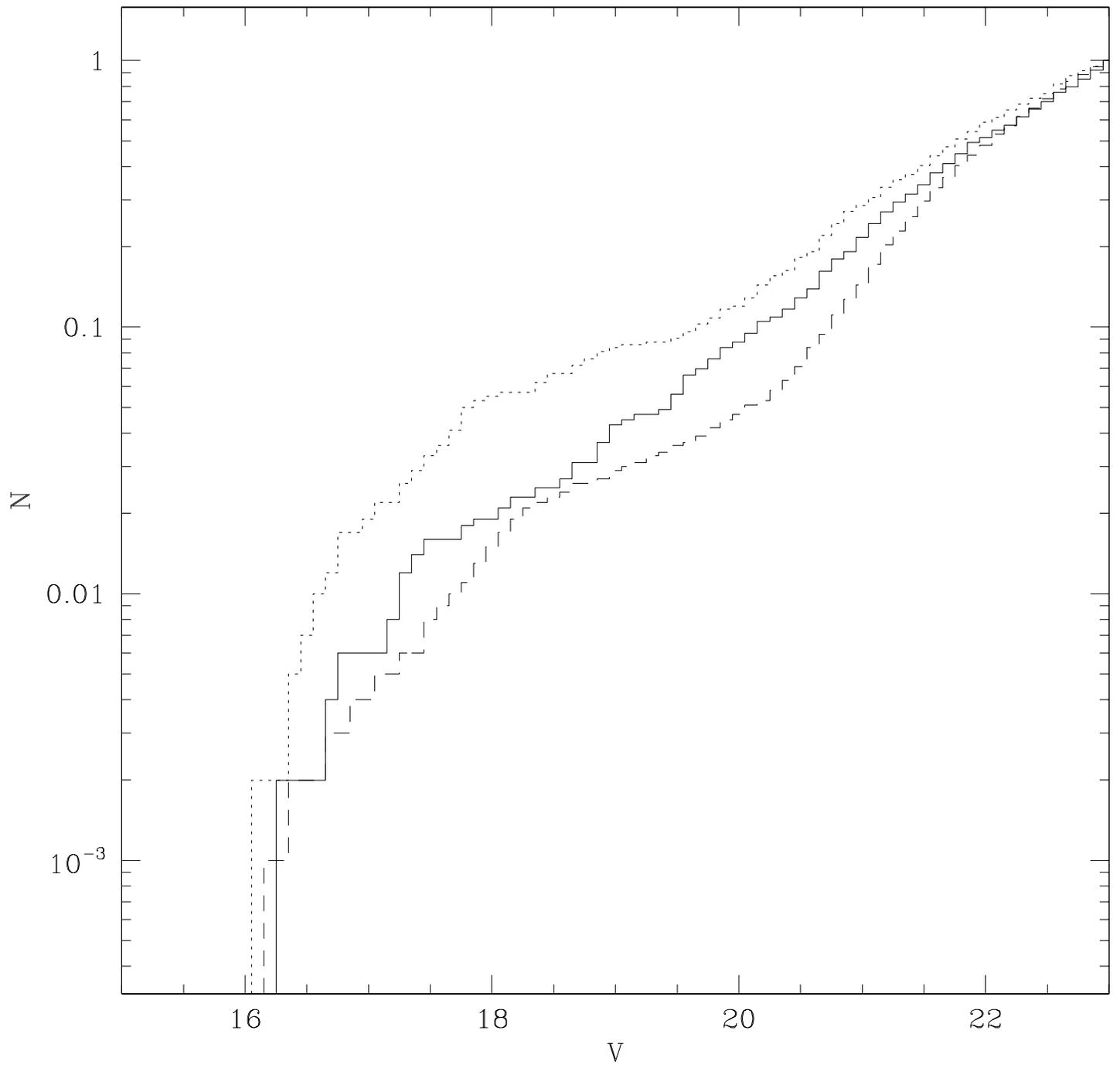
\end{document}